\documentclass[12pt]{article}
\usepackage{amsmath}
\usepackage{graphicx}
\begin{document}
\baselineskip=18 pt
\begin{center}
{\large{\bf Type III Spacetime with Closed Timelike Curves}}
\end{center}

\vspace{0.2cm}

\begin{center}
{\bf Faizuddin Ahmed}\footnote{\bf faizuddinahmed15@gmail.com ; faiz4U.enter@rediffmil.com}\\
{\it Hindustani Kendriya Vidyalaya, Guwahati-05, Assam, India}\\
{\bf Current Affiliation : National Academy Gauripur, Assam, India, 783331}
\end{center}

\vspace{0.5cm}

 \begin{abstract}
We present a cyclic symmetric space-time, admitting closed time-like curves (CTCs) which appear after a certain instant of time, {\it i. e.,} a time-machine space-time. These closed time-like curves evolve from an initai spacelike hypersurface on the plane $z=constant$ in a causally well-behaved manner. The space-time discussed here is free-from curvature singularities and a 4D generalization of the Misner space in curved space-time. The matter field is of pure radiation with a negative Cosmological constant.

\end{abstract}

\section{Introduction}

One of the most intriguing aspects of Einstein’s theory of gravitation is that solutions of Field Equations admit closed time-like curves (CTC). Presence of CTC in a space-time lead to time-travel which violate the causality condition. The first one being the G\"{o}del's space-time \cite{Go} which admit closed time-like curves (CTC) everywhere and an eternal time-machine space-time. There are a considerable number of space-times in literature that admitting closed time-like curves have been constructed. A small sample would be \cite{Go,Sto,Tip,Gott,Bon1,Bon2,Bon3,Kerr,Cart,Gurs,Alcu,lobo,MTY1,MTY2,Kras,Eve,Ori1,Ori2,Ori3,Ori4,Sar}. One way of classifying such causality violating space-times would be to categorize the metrics as either eternal time-machine in which CTC always exist (in this class would be \cite{Go,Sto}), or as time-machine space-times in which CTC appear after a certain instant of time. In the latter category would be the one discussed in \cite{Ori2,Ori3,Ori4}. Many of the models, however, suffer from one or more severe drawbacks. For instance, in some of these solutions, for example \cite{MTY1,MTY2,Ori4} the weak energy condition (WEC) is violated indicating unrealistic matter-energy content and some other solutions have singularities. 

Among the time-machine space-times, we mention two: the first being Ori’s compact core \cite{Ori1} which is represented by a vacuum metric locally isometric to pp waves and second, which is more relevant to the present work, the Misner space \cite{Misn} in 2D. This is essentially a two dimensional metric (hence flat) with peculiar identifications. The Misner space is interesting in the context of CTC as it is a prime example of a space-time where CTC evolve from causally well-behaved initial conditions.

The metric for the Misner space \cite{Misn}
\begin{equation}
ds_{Misn}^2=-2\,dt\,dx-t\,dx^2 \quad .
\label{1}
\end{equation}
where $-\infty<t<\infty$ but the co-ordinate $x$ periodic. The metric (\ref{1}) is regular everywhere as $det g=-1$ including at $t = 0$. The curves $t = t_0$, where $t_0$ is a constant, are closed since $x$ is periodic. The curves $t<0$ are spacelike, but $t>0$ are time-like and the null curves $t=t_0=0$ form the chronology horizon. The second type of curves, namely, $t=t_0>0$ are closed time-like curves (CTC). This metric has been the subject of intense study and quite recently, D. Levanony {\it et al} \cite{Leva} have studied the motion of extended bodies in the 2D Misner space and its flat 4D generalizations. A non-flat 4D space-time, satisfying all the energy conditions, but with causality violating properties of the Misner space, primarily that CTC evolve smoothly from a initially causally well-behaved stage, would be physically more acceptable as a time-machine space-time.

In this paper, we shall attempt to show that causality violating curves appear in non-vacuum space-time with comparatively simple structure. In {\it section} 2, we analyze the space-time, in {\it section} 3, the matter distribution and the energy condition, in {\it section} 4, the space-time is classified and discussed its kinematical properties and concluding one in {\it section} 5.

\section{Analysis of the space-time}

Consider the following metric  
\begin{equation}
ds^2=4\,r^2\,dr^2+e^{2\,\alpha\,r^2}\,\left(dz^2-t\,d\phi^2-2\,dt\,d\phi\right)+4\,\beta\,z\,r\,e^{-\alpha\,r^2}\,dr\,d\phi,
\label{2}
\end{equation}
where $\phi$ coordinate is assumed periodic $0\leq \phi\leq \phi_0$, with $\alpha$ is an integer and $\beta>0$ is real number. We have used co-ordinates $x^1=r$, $x^2=\phi$, $x^3=z$ and $x^{4}=t$. The ranges of the other co-ordinates are $t,z\in(-\infty,\infty)$ and $0\leq r <\infty$. The metric has signature $(+,+,+,-)$ and the determinant of the corresponding metric tensor $g_{\mu\nu}$, $det\;g=-4\,r^2\,e^{6\,\alpha\,r^2}$. The non-zero components of the Einstein tensor are
\begin{eqnarray}
G^{\mu}_{\mu}&=&3\,\alpha^2\nonumber\\
G^{t}_{\phi}&=&-\frac{1}{2}\,e^{-6\,\alpha\,r^2}\,\beta^2
\label{component}
\end{eqnarray}
Consider an azimuthal curves $\gamma$ defined by $r=r_0$, $z=z_0$ and $t=t_0$, where $r_0, z_0, t_0$ are constants, then we have from the metric (\ref{2})
\begin{equation}
ds^2=-t\,e^{2\,\alpha\,r^2}\,d\phi^2
\label{3}
\end{equation}
These curves are null for $t=0$, spacelike throughout for $t=t_0<0$, but become time-like for $t=t_0>0$, which indicates the presence of closed timelike curves (CTC). Hence CTC form at a definite instant of time satisfy $t=t_0>0$.

It is crucial to have analysis that the above CTC evolve from an spacelike $t=constant$ hypersurface (and thus $t$ is a time coordinate)\cite{Ori1}. This can be ascertained by calculating the norm of the vector $\nabla_{\mu} t$ (or by determining the sign of the component $g^{tt}$ in the inverse metric tensor $g^{\mu\nu}$\cite{Ori1}). We find from (\ref{2}) that
\begin{equation}
g^{tt}=t\,e^{-2\,\alpha\,r^2}+\beta^2\,z^{2}\,e^{-6\,\alpha\,r^2}
\label{4}
\end{equation}
A hypersurface $t=constant$ is spacelike provided $g^{tt}<0$ for $t=t_0<0$, but become time-like provided $g^{tt}>0$ for $t=t_0>0$. Here we choose the $z-planes$ defined by $z=z_0$, ($z_0$, a constant equal to zero) such that the above condition is satisfied. Thus the spacelike $t=constant<$0 hypersurface can be chosen as initial conditions over which the initial may be specified. There is a Cauchy horizon for $t=t_0=0$ called Chronology horizon which separates the causal and non-causal of the space-time.  Hence the space-time evolves from a partial Cauchy hypersurface (initial spacelike hypersurface) in a causally well-behaved manner, upto a moment, i.e., a null hypersurface $t=0$ and CTC form at a definite instant of time on $z=constant$ plane.

Consider the Killing vector $\eta=\partial_{\phi}$ for metric (\ref{2}) which has the normal form 
\begin{equation}
\eta^{\mu}=\left (0,1,0,0\right )\quad
\label{5}
\end{equation}
Its co-vector is 
\begin{equation}
\eta_{\mu}=\left(2\,\beta\,z\,r\,e^{-\alpha\,r^2},-t\,e^{2\,\alpha\,r^2},0,-e^{2\,\alpha\,r^2}\right)\quad
\label{6} 
\end{equation}  
(\ref{5}) satisfies the Killing equation $\eta_{\mu\,;\,\nu}+\eta_{\nu\,;\,\mu}=0$. For cyclicly symmetric metric, the norm $\eta_{\mu}\,\eta^{\mu}$ of the Killing vector is spacelike, closed orbits \cite{Marc,Carter,Brane,Bick,Wald}. We note that
\begin{equation}
\eta^{\mu}\,\eta_{\mu}=-t\,e^{2\,\alpha\,r^2}
\label{7}
\end{equation}
which is spacelike for $t<0$, closed orbits ($\phi$ co-ordinate being periodic).

An important note is that the Riemann tensor $R_{\mu\nu\rho\sigma}$ can be expressed in terms of metric tensor $g_{\mu\nu}$ as
\begin{equation}
R_{\mu\nu\rho\sigma}=k\left(g_{\mu\rho}\,g_{\nu\sigma}-g_{\mu\sigma}\,g_{\nu\rho}\right)\quad
\label{8}
\end{equation}
where $k=-\alpha^2$ for the space-time (\ref{2}).

Another important note is that if we take $\beta=0$, then the space-time represented by (\ref{2}) is maximally symmetric vacuum space-time and locally isometric anti-de Sitter space in four-dimension. One can easily show by a number of transformations the standard form of locally isometric $AdS_4$ metric \cite{Zof}
\begin{equation}
ds^2=\frac{3}{(-\Lambda)\,x^2}\left(-dt^2+dx^2+d\phi^2+dz^2\right)\quad
\label{ads}
\end{equation}
where one of the co-ordinate $\phi$ being periodic.

\section{Matter distribution of the space-time and the energy condition}

The Einstein's Field Equations taking into account the cosmological constant
\begin{equation}
G^{\mu\nu}+\Lambda\,g^{\mu\nu}=T^{\mu\nu}\quad,\quad \mu,\nu=1,2,3,4
\label{field}
\end{equation}
Consider the energy-momentum tensor that of pure radiation field \cite{Steph}
\begin{equation}
T^{\mu\nu}=\rho\,n^{\mu}\,n^{\nu}\quad
\label{pure}
\end{equation}
where $n^{\mu}$ is the null vector defined by
\begin{equation}
n^{\mu}=\left(0,0,0,1\right)\quad
\label{null1}
\end{equation}
The non-zero component of the energy-momentum tensor  
\begin{equation}
T^{t}_{\phi}=-\rho\,e^{2\,\alpha\,r^2}
\label{energ}
\end{equation}
Equating Field Equations (\ref{field}) using (\ref{component}) and (\ref{energ}) we get
\begin{eqnarray}
\Lambda&=&-3\,\alpha^2\\
\rho&=&\frac{1}{2}\,\beta^2\,e^{-8\,\alpha\,r^2}\quad,\quad 0\leq r<\infty
\end{eqnarray}
The energy-density of pure radiation or null dust decreases exponentially with $r$ and vanish at $r\rightarrow \pm\infty$. The matter field pure radiation satisfy the energy condition and the energy density $\rho$ is always positive.

\section{Classification and kinematical properties of the space-time}

For classification of the spacetime (\ref{2}), we can construct the following set of null tetrads $(k,l,m,\bar{m})$ as 
\begin{equation}
k_{\mu}=\left (0,1,0,0\right)\quad, 
\label{9}
\end{equation}
\begin{equation}
l_{\mu}=\left(-2\,\beta\,z\,r\,e^{-\alpha\,r^2},\frac{t}{2}\,e^{2\,\alpha\,r^2},0,e^{2\,\alpha\,r^2}\right )\quad,  
\label{10}
\end{equation}
\begin{equation}
m_{\mu}=\frac{1}{\sqrt{2}}\,\left(2\,r,0,i\,e^{\alpha\,r},0\right )\quad, 
\label{11}
\end{equation}
\begin{equation}
\bar{m}_{\mu}=\frac{1}{\sqrt{2}}\,\left(2\,r,0,-i\,e^{\alpha\,r^2},0\right )\quad,  
\label{12}
\end{equation}
where $i=\sqrt{-1}$. The set of null tetrads above are such that the metric tensor for the line element (\ref{2}) can be expressed as
\begin{equation}
g_{\mu \nu}=-k_{\mu}\,l_{\nu}-l_{\mu}\,k_{\nu}+m_{\mu}\,\bar{m}_{\nu}+\bar{m}_{\mu}\,m_{\nu}\quad.
\label{13}
\end{equation}
The vectors (\ref{9})---(\ref{12}) are null vectors and are orthogonal except for $k_{\mu}\,l^{\mu}=-1$ and $m_{\mu}\,{\bar m}^{\mu}=1$. Using this null tetrad above we have calculated the five Weyl scalars 
\begin{eqnarray}
\Psi_3&=&-\frac{i\,\alpha\,\beta\,e^{-2\,\alpha\,r^2}}{2\,\sqrt{2}}\nonumber\\
\Psi_4&=&-\frac{1}{4}\,\beta\,e^{-2\,\alpha\,r^2}\,\left(i+2\,\alpha\,z\,e^{\alpha\,r^2}\right)\quad
\end{eqnarray}
are non-vanishing, while $\Psi_0=\Psi_1=\Psi_3=0$. The spacetime represented by (\ref{2}) is of type III in the Petrov classification scheme. Note that the non-zero Weyl scalars $\Psi_3$ and $\Psi_4$ are finite at $r\rightarrow 0$ and vanish as $r\rightarrow \pm \infty$ indicating asymptotic flatness of the spacetime (\ref{2}). The metric (\ref{2}) is free-from curvature singularities. The curvature invariant known as Kretchsmann scalar is given by
\begin{equation}
R^{\mu\nu\rho\sigma}\,R_{\mu\nu\rho\sigma}=24\,\alpha^4\quad
\label{14}
\end{equation}
and the curvature scalar
\begin{equation}
R=-12\,\alpha^2
\label{15}
\end{equation}
are constant being non-zero.

Using the null tetrad (\ref{9}) we have calculated the {\it Optical} scalars \cite{Steph} the {\it expnasion}, the {\it twist} and the {\it shear} and they are
\begin{eqnarray}
\nonumber
\Theta&=&\frac{1}{2}\,k^{\mu}_{\,;\,\mu}=0\quad\nonumber,\\
\omega^2&=&\frac{1}{2}\,{k_{[\mu\,;\,\nu]}}\,k^{\mu\,;\,\nu}=0\quad\nonumber,\\
\sigma\,\bar{\sigma}&=&\frac{1}{2}\,{k_{(\mu\,;\,\nu)}}\,k^{\mu\,;\,\nu}-\Theta^2=0\quad
\end{eqnarray}
and the null vector (\ref{9}) satisfy the geodesics equation
\begin{equation}
k_{\mu\,;\,\nu}\,k^{\nu}=0\quad.
\label{16}
\end{equation}
Thus the spacetime represented by (\ref{2}) is non-diverging, have shear-free null geodesics congruence. One can easily show that for constant $r$ and $z$, the metric (\ref{2}) reduces to conformal Misner space in 2D
\begin{equation}
ds_{confo}^2=\Omega\,ds_{Misn}^2\quad
\label{17}
\end{equation}
where $\Omega=e^{-2\,\alpha\,r^2}$ is a constant.

\section{Conclusion}

Our primary motivation in this paper is to write down a metric for a spacetime that incorporates the Misner space and its causality violating properties and to classify it. The solution presented here is non-vacuum, cyclicly symmetric metric (\ref{2}) and serves as a model of time-machine spacetime in the sense that CTC appear at a definite instant of time on the $z-plane$. Most of the CTC spacetimes violate one or more energy conditions or unrealistic matter souce and are unphysical. The model discussed here is free-from all these problems and matter distribution is of pure radiation field with negative cosmological constant satisfying the energy condition.

\begin{center}
\section*{Appendix}
\end{center}

The above metric (\ref{2}) can be transform to a simple form as follows. Let us perform the following transformation
\begin{equation}
r \to \sqrt{r'},
\label{A1}
\end{equation}
into the metric (\ref{2}), one will get
\begin{equation}
ds^2=dr'^2+e^{2\,\alpha\,r'}\,(dz^2-t\,d\phi^2-2\,dt\,d\phi)+2\,\beta\,z\,e^{-\alpha\,r'}\,dr'\,d\phi.
\label{A2}
\end{equation}
Finally doing another transformation 
\begin{equation}
r' \to \frac{1}{\alpha}\,\mbox{ln} \varrho
\label{A7}
\end{equation}
into the metric (\ref{A2}), we get the following line element
\begin{equation}
ds^2=\frac{d\varrho^2}{\alpha^2\,\varrho^2}+\varrho^2\,dz^2+\varrho^2\,(-t\,d\phi^2-2\,dt\,d\phi)+\frac{2\,\beta\,z}{\alpha\,\varrho^2}\,d\varrho\,d\phi.
\label{A8}
\end{equation}

Or if one does the following transformation
\begin{equation}
r' \to \frac{1}{\alpha}\,\mbox{ln} (\alpha\,\varrho)
\label{A3}
\end{equation}
into the metric (\ref{A2}), we get the following line element
\begin{equation}
ds^2=\frac{d\varrho^2}{\alpha^2\,\varrho^2}+\alpha^2\,\varrho^2\,dz^2+\alpha^2\,\varrho^2\,(-t\,d\phi^2-2\,dt\,d\phi)+\frac{2\,\beta\,z}{\alpha^2\,\varrho^2}\,d\varrho\,d\phi.
\label{A4}
\end{equation}

Or if one does the following transformation
\begin{equation}
r' \to -\frac{1}{\alpha}\,\mbox{ln} (\alpha\,\varrho)
\label{A5}
\end{equation}
into the metric (\ref{A2}), we get the following line element
\begin{equation}
ds^2=\frac{1}{\alpha^2\,\varrho^2}\,(d\varrho^2-t\,d\phi^2-2\,dt\,d\phi+dz^2)-2\,\beta\,z\,d\varrho\,d\phi.
\label{A6}
\end{equation}

\end{document}